\def \be {\begin{equation}} 
\def \ee {\end{equation}} 
\def \bea {\begin{eqnarray}} 
\def \eea {\end{eqnarray}} 
\def \bse {\begin{subequations}} 
\def \ese {\end{subequations}} 
\def \bde {\begin{description}} 
\def \ede {\end{description}}

\def \del {\partial} 
\def \dels {\partial\kern-.5em / \kern.5em} 
\def \As {{A\kern-.5em / \kern.5em}} 
\def \Ds {D\kern-.7em / \kern.5em} 

\def\e{\,{\rm e}}

\def \G {\Gamma} 
\def \d {\delta} 
\def \D {\Delta}
 
\def \m {\mu} 
 
\def \k {\kappa} 
\def \lam {\lambda} 
\def \Lam {\Lambda}

\def \r {\rho} 
\def \om {\omega} 
\def \Om {\Omega} 
 
\def \th {\theta}

\def \vp {\varphi}

\def\pl#1{{\sl Phys.~Lett.~\bf B#1}}

\renewcommand{\v}[1]{\boldsymbol{#1}} 
 
\documentclass{article} 
\usepackage{epsf}
\usepackage{graphicx}
\usepackage{threeparttable}
\usepackage{amsmath} 
 
\begin{document} 

 
\begin{center} 

\vskip .5in 
 
\textbf{\large 
Quasinormal Modes of Kerr Black Holes in Four and Higher Dimensions} 
 
\vskip .5in 
{\large
Hsien-chung Kao\footnote{hckao@phy.ntnu.edu.tw} and Dan Tomino\footnote{dan@home.phy.ntnu.edu.tw}
\vskip 15pt

{Department of Physics, National Taiwan Normal University, Taipei, Taiwan 116.}\\ 
}
  
\vskip .2in 
 
\vspace{60pt} 
\end{center} 
\begin{abstract} 
  
We analytically calculate to leading order the asymptotic form of quasinormal frequencies of Kerr black holes in four, five and seven dimensions. All the relevant quantities can be explicitly expressed in terms of elliptical integrals. In four dimensions, we confirm the results obtained by Keshest and Hod by comparing the analytic results to the numerical ones.
  
\end{abstract} 
 
\setcounter{footnote}{0} 
\newpage 
 
\section{Introduction} 
 
Perturbation of black holes are known to reveal characteristic damped oscillation modes which dominate the time evolution in certain intermediate period of time \cite{Vish}.  Since the frequencies are  complex, they are called quasinormal modes (QNMs).  For a general review and classification, see Refs. \cite{NollertReview,Ishibashi,Schiappa}.  They depend only on the fundamental parameters of the black holes, such as mass, angular momentum and charge. For a Schwarzschild black hole, an asymptotic formula for high overtones has been obtained using the monodromy matching method \cite{Motl}:
\be
\om_n = \frac{1}{8\pi M} \biggl\{ \ln 3+ \left(2n- 1\right)\pi i \biggr\} 
\ee 
in the units $G=c=\hbar=1.$  The analytic value $\ln3$ in the real part of the above formula was used to argue that the relevant gauge group in loop quantum gravity should be $SO(3)$ \cite{Hod}. Although this result turns out to be not universal \cite{Motl}, one still expects QNMs to play an important role in understanding black holes and quantum gravity. 

The monodromy method has been generalized so that first order correction to the asymptotic form of quasinormal frequencies can also be calculated \cite{Brink}. In the case of Schwarzschild black holes, comparisons with numerical results have been made and the agreement is excellent \cite{Nollert,Cardoso1}. In Ref. \cite{2ndorder}, the calculation is further extended to the second order.  The agreement is not as good as the first order case. In fact, there seems to be sizable discrepancy in higher angular momentum case ($l=6$).  More study is needed to clarify the situation.

Although high overtones of QNMs in spherically symmetric black holes have been extensively studied,  they are not as well studied in the case of rotating black holes. In Ref. \cite{Cardoso2}, convergent numerical results have been presented. Making use of the monodromy analysis used in Refs. \cite{Motl,Andersson}, Keshet and Hod obtain the first analytic results that are in excellent agreement with the numerical ones \cite{Keshet1}. There are two  distinct properties of rotating black holes that makes its highly damped QNMs quite different from that of spherically symmetric ones. First, there are more than one turning points. Second, there is a term linear in the frequency $\om$, which contributes to the real part in leading order.  In this paper, we will extend their method to find the asymptotic formula of quasinormal frequency for higher dimensional Kerr black holes.

\section{Calculation of the asymptotic form of quasinormal frequencies}

Consider a massless scaler field $\Phi$ in the background of a $D$-dimensional Kerr black hole. Using the Boyer-Linquist coordinate, we can write the metric as
\bea
ds^2 = -\frac{\D}{\r^2}\bigl[dt - a\sin\th^2 d\phi \bigr]^2 + \r^2\left[\frac{dr^2}{\D} + d\th^2 \right] + \frac{\sin^2\th}{\r^2}\left[ adt - (r^2 + a^2)d\phi\right]^2 + r^2\cos^2\th d\Om^2_{D-4}.
\eea 
Here,
\bea
&\;& \r^2= r^2 +a^2 \cos^2\th, \\
&\;& \D= r^2 -\m\, r^{5-D} +a^2.
\eea
$d\Om^2_{D-4}$ is the metric on $S_{D-4}/Z_2$, with $\Om_{D-4} = (\psi_1, \ldots, \psi_{D-4})$ and $0\le \psi_i \le \pi$ for all $i$. The location of an event horizon is determined by $\D=0$.  For $D=4$, there are both the outer and inner horizons $r_\pm = \m/2 \pm \sqrt{\m^2/4 - a^2}$.  For $D\ge 5$, there is only one horizon.  The surface $r= r_+$ is also a Killing horizon. The corresponding Killing vector is $\v{\xi} = \v{\del}_t + \Om_H \v{\del}_\phi$. Here, $\Om_H = \frac{a}{r_+^2 + a^2}$ is the angular velocity on the horizon. From $\v{\xi}$, one can find surface gravity $\k = \frac{(D-3)r_+^2 + (D-5)a^2}{2r_+ (r_+^2 + a^2)}$.   The ADM mass is given by $M = \frac{(D-2) {\cal A}_{D-2}\, \m}{16\pi}$, with ${\cal A}_n = \frac{2 \pi^{(n+1)/2}}{\G(\frac{n+1}{2})}$ the area of an $n$-dimensional unit sphere.  The Bekenstein-Hawking entropy is given by $S = \frac{1}{4}{\cal A}_{D-2}(r_+^2 + a^2)r_+^{D-4}$ \cite{Bekenstein1}. Making use of the identity $dS = \frac{2\pi}{\k}(dM - \Om_H dJ)$, one then find $J= \frac{{\cal A}_{D-2}(r_+^2 + a^2)a}{8\pi}$.

It's well known that $\Phi$ satisfies the Klein-Gordon equation:
\begin{eqnarray}
{1 \over \sqrt{-g}}\partial_{\mu}\left\{g^{\mu \nu}\sqrt{-g}\partial_{\nu}\Phi \right\}=0.
\end{eqnarray}
Let 
\be
\Phi(r,t,\th,\phi,\Om_{D-4})= \tilde{R}_{lmL}(r) S_{lmL}(u) Y_{L,M_1,\ldots,M_{D-5}}(\Om_{D-4}) \e^{-i\om t + im\phi }, \label{Fourier}
\ee
where $u=\cos^2\th$ and $Y_{L,M_1,\ldots,M_{D-5}}(\Om_{D-4})$ the spherical harmonics.
$R_{lmL}(r)$ and $S_{lmL}(u)$ satisfy
\bea
&\;& \hskip -2cm \frac{1}{r^{D-4}} \frac{d }{dr}  \left[r^{D-4}\D \frac{d R_{lmL}(r)}{dr}\right]\nonumber \\
&\;& \hskip -2cm + \left\{\frac{\om^2[(r^2+a^2)^2-a^2\D]-\om[2m\m a]+a^2[m^2-L(L+D-5)\D/r^2]}{r^{D-5}\D} - \Lam_{lmL}\right\} R_{lmL}(x) =0, \\
&\;& \hskip-2cm u(1-u)\frac{d^2 S_{lmL}(u)}{du^2} + \frac{1}{2}[(D-3)-(D-1)u]\frac{d S_{lmL}(u)}{du} \nonumber \\
&\;& \hskip-2cm + \frac{1}{4}\left[\om^2 a^2 u - \frac{m}{1-u} - \frac{L(L+D-5)}{u} + \Lam_{lmL}\right] S_{lmL}(u) = 0,
\eea
respectively.  $\Lam_{lmL}$ is the separation constant.  Define $T_{lmL}(u)= u^{-L/2}(1-u)^{-m/2}S_{lmL}(u)$ and the angular equation becomes
\bea
&\;& \hskip-4cm u(1-u)\frac{d^2 T_{lmL}(u)}{du^2} + \left[ (\frac{D-3}{2} + L) - (\frac{D-1}{2} +  m + L)\, u \right]\frac{d T_{lmL}(u)}{du} \nonumber \\
&\;& \hskip-4cm + \frac{1}{4}\left[\om^2 a^2 u - (m+L)(m+L+D-3) + \Lam_{lmL}\right] T_{lmL}(u) = 0.
\eea
When $a \to 0$, the solution to the above equation becomes the hypergeometric function $ _2 F_1(A,B;C,u)$ with 
\bea
&\;& \hskip -7cm A = \frac{1}{2}\left[m + L + \frac{D-3}{2} - \sqrt{ \left(\frac{D-3}{2} \right)^2 + \Lam_{lmL}}\, \right], \\
&\;& \hskip -7cm B = \frac{1}{2}\left[m + L + \frac{D-3}{2} + \sqrt{ \left(\frac{D-3}{2} \right)^2 + \Lam_{lmL}}\, \right], \\
&\;& \hskip -7cm C = L + \frac{D-3}{2}.
\eea
In such limit, $T^{(0)}_{lmL}(u)$ together with $\e^{im\phi}$ and $Y_{L,M_1,\ldots,M_{D-5}}(\Om_{D-4})$ form the spherical harmonics on $S^{D-2}$, and therefore $\Lam_{lmL} = l(l+D-3)$.  As a result, $A = \frac{1}{2}(m+L-l)=-p$, where $p$ specifies the number of zeros of $T^{(0)}_{lmL}(u)$ for $u$ in the interval $[0,1]$.

It is straightforward to generalize Flammer's method to arbitrary dimension $D$ \cite{Flammer}.  In the limit $\om\to -i \infty$,
\bea
&\;& \Lam_{lmL} = \Lam_0 c + \Lam_1 + O (c^{-1}),  \\
&\;& T_{lmL}(u) = Z_0(x)+\frac{1}{c}Z_1(x) + O(c^{-2}),
\eea
where $c= i\om a$ and $x=cu$. To leading order in $c^{-1}$, $Z_0(x)$ satisfies 
\bea
&\;& x Z''_0(x) + \frac{1}{2}(D-3+2L)Z'_0(x)-\frac{1}{4}(x-\Lam_0)Z_0(x) =0.
\eea
The solution is given by $Z_0(x)=\e^{-x/2}L_{\frac{1}{4}(\Lam_0-D+3-2L)}^{(L)}(x)$, where $L^{(n)}_p(x)$ is the generalized Laguerre polynomials with $p$ zeros. For $Z_0(x)$ to have the same number of zeros as $T^{(0)}_{lmL}(u)$, we find
\bea
\Lam_0 =[2(l-|m|)+(D-3)].
\eea
This is consistent with the result in the $D=4$ case \cite{Cardoso2}.
Let $R_{lmL}(r) = [r^{D-4}\D]^{-1/2} \tilde{R}_{lmL}(r)$, and the radial equation becomes 
\bea
&\;& \hskip -2cm  \left[\frac{d^2 }{dr^2}  + \frac{\om^2 q_0(r)+ \om q_1(r)+ q_2(r)}{\D^2} \right]\tilde{R}_{lmL}(r) =0. \label{RadialEq1}
\eea
Here,
\bea
&\;& \hskip -2cm q_0(r)= (r^2+a^2)^2-a^2\D = r^4 + a^2 r^2 + \m a^2 r^{5-D}; \\
&\;& \hskip -2cm q_1(r)= m[-2\m a r^{5-D}] + \Lam_0 [-ia\D]; \\
&\;& \hskip -2cm q_2(r)= m^2[a^2] + L(L+D-5)[\frac{-a^2\D}{r^2}]+\Lam_1[-\D] \nonumber \\
&\;& \hskip -1cm - \frac{(D-2)(D-4)\D^2(r)-4(D-5)a^2\D-[2a^2-(D-3)\m r^{5-D}]^2}{4r^2}.
\eea
Similar to Ref.\cite{Keshet1}, we define $z\equiv \int^r V(r') dr'$ with $V(r)= \D^{-1}(r)[q_0(r)+\om^{-1}q_1(r)]^{1/2}$.  The radial equation becomes
\bea
\left\{-\frac{d^2}{dz^2} -\om^2 - V_1(r) \right\}\hat{R}_{lmL}(r)=0,
\eea
where $\hat{R}= V^{1/2} \tilde{R}$ and
\bea
V_1(r)= -\frac{q_2(r)}{\D^2(r)V^2(r)} + \frac{V''(r)}{2V^3(r)} - \frac{3[V'(r)]^2}{4V^4(r)}.
\eea
It has been shown using the monodromy matching method that the condition for quasinormal modes is given by
\bea
2i\om \int_{C_{t,o}} V dr =2\pi i\left(n + \frac{1}{2} \right). \label{QNMCond}
\eea
Here, $C_{t,o}$ is a contour in the complex plane that runs from $t_1$ to $t_2$ lying beyond the outer horizon $r= r_+$.  $t_1, t_2$ are the transition points in the right half complex plane determined by the condition:
\bea
q_0(t)+\om^{-1} q_1(t) = 0.
\eea

\vspace*{-0.5cm}
\begin{figure}[tbh]
\centerline{\epsfxsize=7cm \epsffile{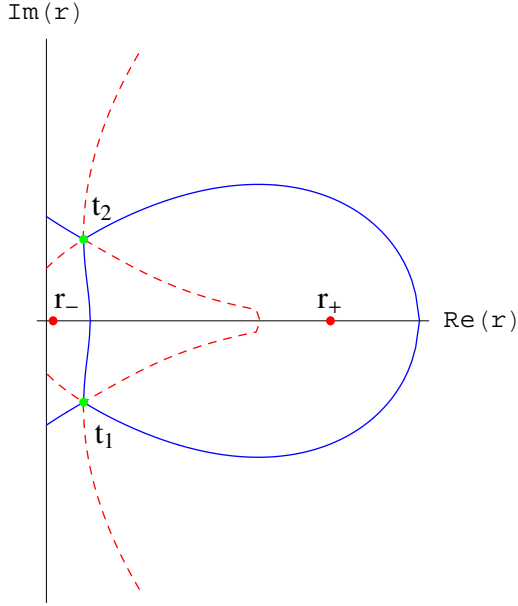}}
\caption{\label{fig:stokeline} The Stokes lines (dashed) and anti-Stokes lines (solid) emanting from the turning points $t_1$ and $t_2$ in complex $r$-plane for $D=4$ $a=0.5$ are shown.  $r_-$ and $r_+$  are the inner and outer horizon radii, repectively.}
\end{figure}

In the limit $\om \to -i\infty$,
\bea
2\om \int_{C_{t,o}} V dr \to 2\om \int_{C_{r,o}} \frac{\sqrt{q_0(r)}}{\D} dr + \int_{C_{r,o}} \frac{q_1(r)}{\D\sqrt{q_0(r)}} dr.
\eea 
Making use of the explicit expression for $q_1(r)$ and following the convention in Ref. \cite{Keshet1}, we have
\bea
&\;& \hskip -9cm \d_0 = -2i\int_{C_{r,o}} \frac{\sqrt{q_0(r)}}{\D} dr;  \label{delta0}\\
&\;& \hskip -9cm \d_m = 2i\int_{C_{r,o}} \frac{\m a r^{5-D}}{\D\sqrt{q_0(r)}} dr; \label{deltam}\\
&\;& \hskip -9cm \d_\Lam = 2i\int_{C_{r,o}} \frac{a}{2\sqrt{q_0(r)}} dr. \label{deltalambda}
\eea
The sign convention is opposite to that in Ref. \cite{Keshet1}. It is chosen to make $\d_0, \d_m$ and $\d_\Lam$ all positive. The QNM quantization condition in eq (\ref{QNMCond}) becomes
\bea
\om \d_0 + m \d_m + i \Lam_0 \d_\Lam = -i\pi(2n+1).
\eea
To leading order in $c^{-1}$,
\bea
\om  = -i(n\hat{\d} +\hat{\phi}) - m \hat{\om},
\eea
where $\hat{\d} = 2\pi/\d_0$, $\hat{\phi} = (\Lam_0 \d_\Lam +\pi)/\d_0$, $\hat{\om} = \d_m/\d_0$.
So far, our results have been completely general and hold for arbitrary dimensions.  It has been pointed out that the integrals for $\d_0, \d_m$ can be expressed in terms of elliptical functions for $D=4$ \cite{Keshet1}. From eqs (\ref{delta0}) to (\ref{deltalambda}), it is easy to see that similar situation also occurs for $D=5$ and $7$. Therefore, we will only focus on these cases hereafter.

Since no explicit expressions for the $D=4$ case are given in Ref. \cite{Keshet1}, we think it is appropriate to present them here. The explicit form of $q_0(r)$ is given by
\bea
q_0(r)=r^4 + a^2 r^2 + \m a^2 r.
\eea
There are four transition points. To leading order in $c^{-1}$, they are $r=0$, $r= -2u_1$, and $r = u_1 \pm i v_1$, with 
\bea
&\;& \hskip -9cm u_1 = \frac{a(\lam^{1/3}- \lam^{-1/3})}{2\sqrt{3}}, \\
&\;& \hskip -9cm v_1 = \frac{a(\lam^{1/3}+ \lam^{-1/3})}{2},  \\
&\;& \hskip -9cm \lam = \frac{3\sqrt{3} \m}{2a} + \sqrt{1+ \frac{27\m^2}{4a^2}}.
\eea
For convenience, define 
\bea
&\;& f_0(r_0,u_1,v_1) = \int_{u_1-iv_1}^{u_1+iv_1}  \frac{\sqrt{r(r+2u_1)[(r-u_1)^2+v_1^2]}}{(r-r_0)}dr; \\
&\;& f_m(r_0,u_1,v_1) = \int_{u_1-iv_1}^{u_1+iv_1} \frac{1}{(r-r_0)\sqrt{r(r+2u_1)[(r-u_1)^2+v_1^2]}} dr.
\eea
Using mathematica, we find 
\bea
&\;& \hskip -1.5cm f_0(r_0,u_1,v_1) = i r_0 \sqrt{3u_1^2 + v_1^2 + 2i u_1 v_1} E\left[\frac{4i u_1 v_1}{3u_1^2+v_1^2+2i u_1 v_1}\right] - \frac{i (r_0-u_1)(9u_1^2+v_1^2)}{\sqrt{3u_1^2+v_1^2+2i u_1 v_1}} K\left[\frac{4i u_1 v_1}{3u_1^2+v_1^2+2i u_1 v_1}\right] \nonumber \\
&\;& \hskip 0.8cm - \frac{i2r_0(3u_1 + iv_1) (r_0-u_1+iv_1)}{\sqrt{3u_1^2+v_1^2+2i u_1 v_1}} \Pi\left[\frac{-2i v_1(r_0+2u_1)}{(r_0-u_1-iv_1)(3u_1-iv_1)},\frac{4i u_1 v_1}{3u_1^2+v_1^2+2i u_1 v_1}\right] \nonumber \\
&\;& \hskip 0.8cm - \frac{i(3u_1+iv_1)(3u_1^2-v_1^2-2r_0^2)}{\sqrt{3u_1^2+v_1^2+2i u_1 v_1}} \Pi\left[\frac{-2iv_1}{3u_1-iv_1}, \frac{4i u_1 v_1}{3u_1^2+v_1^2+2i u_1 v_1}\right], \label{f0D4}
\eea
and
\bea
&\;& \hskip -2cm f_m(r_0,u_1,v_1) = \frac{2}{r_0 \sqrt{3u_1^2 + v_1^2 - 2i u_1 v_1}} \nonumber \\
&\;& \hskip 0.3cm \times \Biggl\{ F\left[\sin^{-1}\left(\sqrt{\frac{3u_1^2 + v_1^2 - 2i u_1 v_1}{3u_1^2 + v_1^2 + 2i u_1 v_1}}\,\right),\frac{3u_1^2 + v_1^2 + 2i u_1 v_1}{3u_1^2 + v_1^2 - 2i u_1 v_1} \right] - K\left[\frac{3u_1^2 + v_1^2 + 2i u_1 v_1}{3u_1^2 + v_1^2 - 2i u_1 v_1} \right]  \Biggr\} \nonumber \\
&\;& \hskip 0.3cm - \frac{4u_1}{r_0(r_0+2u_1) \sqrt{3u_1^2 + v_1^2 - 2i u_1 v_1}} \nonumber \\
&\;& \hskip 0.3cm \times \Biggl\{ \Pi\left[\frac{r_0(3u_1-i v_1)}{(r_0+2u_1)(u_1-iv_1)}, \sin^{-1}\left(\sqrt{\frac{3u_1^2 + v_1^2 - 2i u_1 v_1}{3u_1^2 + v_1^2 + 2i u_1 v_1}}\, \right), \frac{3u_1^2 + v_1^2 + 2i u_1 v_1}{3u_1^2 + v_1^2 - 2i u_1 v_1} \right] \nonumber \\
&\;& \hskip 0.8cm - \Pi\left[\frac{r_0(3u_1-i v_1)}{(r_0+2u_1)(u_1-iv_1)},\frac{3u_1^2 + v_1^2 + 2i u_1 v_1}{3u_1^2 + v_1^2 - 2i u_1 v_1} \right] \Biggr\}. \label{fmD4}
\eea
Here, $E(m), E(\vp,m), K(m), F(\vp,m), \Pi(n,m)$ and $\Pi(n,\vp,m)$ are the elliptical integrals:
\bea
&\;& E(m) = \int_0^{\frac{\pi}{2}} [1-m\sin^2\th]^{1/2} \,d\th, \\
&\;& E(\vp,m) = \int_0^{\vp} [1-m\sin^2\th]^{1/2} \,d\th, \\
&\;& K(m) = \int_0^{\frac{\pi}{2}} [1-m\sin^2\th]^{-1/2} \,d\th, \\
&\;& F(\vp,m) = \int_0^{\vp} [1-m\sin^2\th]^{-1/2} \,d\th, \\
&\;& \Pi(n,m) = \int_0^{\frac{\pi}{2}} [1-n\sin^2\th]^{-1}[1-m\sin^2\th]^{-1/2} \,d\th, \\
&\;& \Pi(n,\vp,m) = \int_0^{\vp} [1-n\sin^2\th]^{-1}[1-m\sin^2\th]^{-1/2} \,d\th. 
\eea 
Note that because of the pole at $r=r_0$ and branch cuts, there is ambiguity in eqs (\ref{f0D4}) and (\ref{fmD4}).   To find the correct analytic expression, we compare the result with the numerical integration.  After some trial and error, we find
\bea
&\;& \hskip -4cm \d_0 = -2i\biggl\{ \frac{f_0(r_+,u_1,v_1) - f_0(r_-,u_1,v_1)}{r_+ - r_-} + \frac{i2\pi \m r_+}{r_+ - r_-} \th(3u_1 r_+ - 3 u_1^2 - v_1^2) \biggr\},
\eea
where $\th(x)$ is the step function.  The term with a step function is introduced to compensate the discontinuity caused by the term $\Pi\left[\frac{-2i v_1(r_+ + 2u_1)}{(r_+ - u_1 - iv_1)(3u_1-iv_1)},\frac{4i u_1 v_1}{3u_1^2+v_1^2+2i u_1 v_1}\right]$.  The magnitude of jump is given by the residue of $q_0^{1/2}/\D$ at $r=r_+$.
Similarly,
\bea
&\;& \hskip -6.8cm \d_m = -2i \biggl\{ \frac{-\m a[r_+ f_m(r_+,u_1,v_1) - r_- f_m(r_-,u_1,v_1)]}{r_+ - r_-} \biggr\}.
\eea
Note that here there is no discontinuity and we do not need to introduce any term with step function.  Comparisons to the numerical results from Ref. \cite{Cardoso2} are given in Fig.2 and 3.

\vspace*{0.5cm}
\begin{figure}[tbh]
\centerline{\epsfxsize=8cm \epsffile{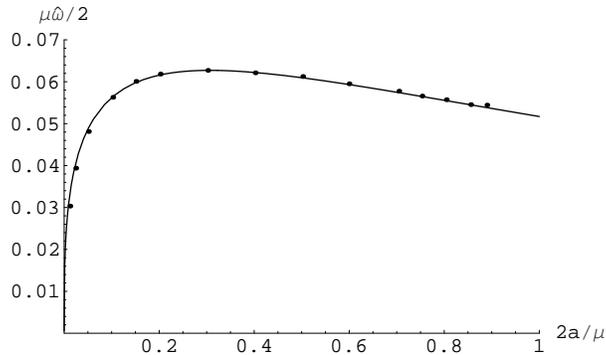}}
\caption{\label{fig:omega_hat_D=4} Comparison between the analytic and numerical results for the real part of the highly damped QNM frequency $\hat{\om}(a)$.}
\end{figure}

\vspace*{0.5cm}
\begin{figure}[tbh]
\centerline{\epsfxsize=8cm \epsffile{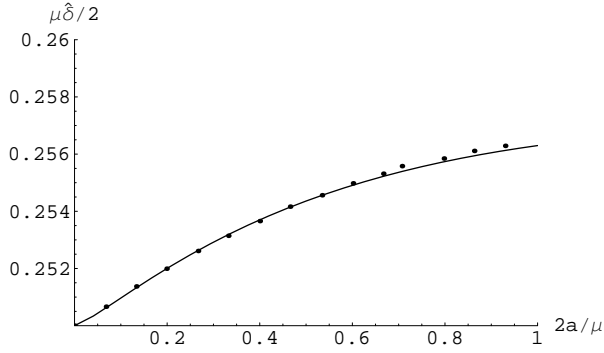}}
\caption{\label{fig:delta_hat_D=4} Comparison between the analytic and numerical results for the level spacing $\hat{\d}$.}
\end{figure}

In the extremal limit, $a\to \m/2$. As a result, $\lam \to \frac{\sqrt{7}+\sqrt{3}}{2},$$u_1 \to \frac{\m}{4}$,  and $v_1 \to \frac{\sqrt{7}\m}{4}$.  Using L'Hospital's rule, we find
\bea
&\;& \hskip -2.5cm \d_0 = \frac{i\sqrt{1-i\sqrt{7}}}{2}\Biggl\{ 2(3-i\sqrt{7})E \biggl[\frac{7+i 5\sqrt{7}}{16}\biggr] - 3(1-i\sqrt{7})K \biggl[\frac{7+i 5\sqrt{7}}{16}\biggr] \nonumber \\
&\;& - 2(3 + i\sqrt{7})\Pi \biggl[\frac{7+i \sqrt{7}}{4},\frac{7+i 5\sqrt{7}}{16}\biggr]
+ 2(5 - i\sqrt{7})\Pi \biggl[\frac{7 - i 3\sqrt{7}}{8},\frac{7+i 5\sqrt{7}}{16}\biggr] \Biggr\}  \approx 24.52, \\
&\;& \hskip -2.5cm \d_m =  \frac{i\sqrt{-5+i\sqrt{7}}}{8}\Biggl\{ (5 + i\sqrt{7}) E \biggl[\frac{7+i 5\sqrt{7}}{16}\biggr] - 8 K \biggl[\frac{7+i 5\sqrt{7}}{16}\biggr]  \Biggr\} \approx 1.27.
\eea
The limit is smooth in constrast to the Reissner-Nordstrom case as pointed out in Ref. \cite{Keshet1}.  In the Schwarzschild limit $a \to 0$, even though the asymptotic QNMs are not continuous the level spacing does goes to the Schwarzschild result $\hat{\d} = (D-3)/(2\m)$ smoothly.

For $D=5$, 
\bea
q_0(r)=r^4 + a^2 r^2 + \m a^2.
\eea
There are four transition points:  $r = \pm u_1 \pm i v_1$, with 
\bea
&\;& \hskip -8cm u_1 = a\sqrt{\frac{\sqrt{\m}}{2a}- \frac{1}{4}}, \\
&\;& \hskip -8cm v_1 = a\sqrt{\frac{\sqrt{\m}}{2a} + \frac{1}{4}},  \\
&\;& \hskip -8cm r_+ = a \sqrt{\frac{\m}{a^2}-1}.
\eea
Here, we have
\bea
&\;& \hskip -2.7cm \d_0 =
 \left(\frac{4\pi \m}{2 r_+} \right) + 2i(u_1+ i v_1)\left\{E\left[\sin^{-1}\left( \frac{u_1 + i v_1}{u_1 - i v_1} \right),\frac{(u_1 -i v_1)^2}{(u_1 +i  v_1)^2}\right]-E\left[\frac{(u_1 -i v_1)^2}{(u_1 +i  v_1)^2}\right] \right\} \nonumber \\
&\;& \hskip -2.0cm 
- \frac{2i\left[r_+^2 - (u_1 - i v_1)^2 \right]}{(u_1+ i v_1)}\left\{F\left[\sin^{-1}\left( \frac{u_1 + i v_1}{u_1 - i v_1} \right),\frac{(u_1 -i v_1)^2}{(u_1 +i  v_1)^2}\right]-K\left[\frac{(u_1 -i v_1)^2}{(u_1 +i  v_1)^2}\right] \right\} \nonumber \\
&\;& \hskip -2.0cm 
+ \frac{2i\m^2}{r_+^2(u_1+ i v_1)}\left\{\Pi\left[\frac{(u_1 - i v_1)^2}{r_+^2}, \sin^{-1}\left( \frac{u_1 + i v_1}{u_1 - i v_1} \right),\frac{(u_1 -i v_1)^2}{(u_1 +i  v_1)^2}\right]-\Pi\left[\frac{(u_1 - i v_1)^2}{r_+^2}, \frac{(u_1 -i v_1)^2}{(u_1 +i  v_1)^2}\right] \right\}.
\eea
\bea
&\;& \hskip -3.8cm \d_m = \left(\frac{-4\pi a}{2 r_+} \right) \nonumber \\
&\;& \hskip -3.1cm - \frac{2i\sqrt{\m}(u_1 - i v_1)}{r_+^2} \left\{ \Pi\left[\frac{(u_1 - i v_1)^2}{r_+^2}, \sin^{-1}\left( \frac{u_1 + i v_1}{u_1 - i v_1} \right),\frac{(u_1 -i v_1)^2}{(u_1 +i  v_1)^2}\right] - \Pi\left[\frac{(u_1 - i v_1)^2}{r_+^2}, \frac{(u_1 -i v_1)^2}{(u_1 +i  v_1)^2}\right] \right\}.
\eea
The results are shown in Fig. 4 and 5.  In contrst to the $D=4$ case, there is a critical value $a_c \approx 0.678\m^{1/2}$. When $a>a_c$, $\hat{\om}$ changes sign.  The physical meaning is not clear at the moment.  It might be related to the "algebraically special" frequencies \cite{Chandrasekhar}.

\begin{figure}[tbh]
\centerline{\epsfxsize=8cm \epsffile{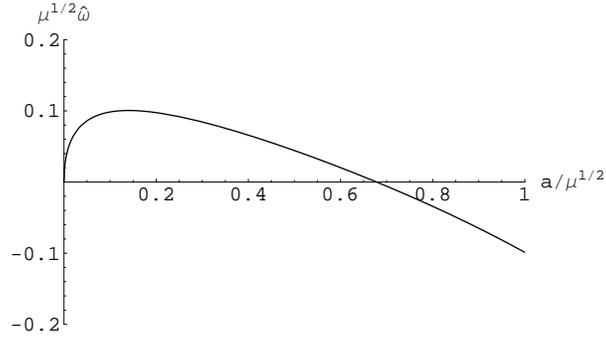}}
\caption{\label{fig:omega_hat_D=5} The analytic result for the real part of the highly damped QNM frequency $\hat{\om}(a)$ when $D=5$.}
\end{figure}

\begin{figure}[tbh]
\centerline{\epsfxsize=8cm \epsffile{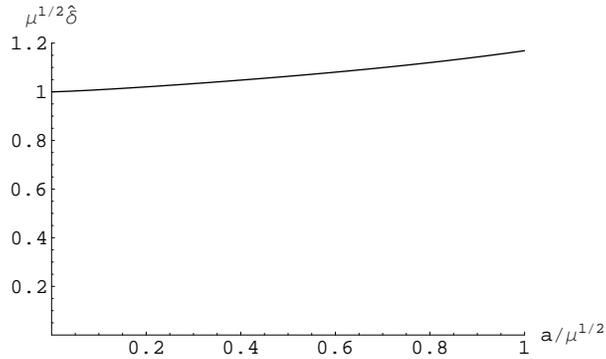}}
\caption{\label{fig:delta_hat_D=5} Analytic result for level spacing $\hat{\d}$ when $D=5$.}
\end{figure}

For $D=7$, 
\bea
q_0(r)=r^4 + a^2 r^2 + \m a^2 r^{-2}.
\eea
There are six transition points: $r = \pm u_1 \pm i v_1$ and $\pm i w_1$. It is convenient to introduce the variable $y= r^2$. Consequently,
\bea
&\;& \hskip -1cm \int_{u_1-iv_1}^{u_1+iv_1} \frac{r\sqrt{r^6+ a^2 r^4 + \m a^2}}{r^4 + a^2 r^2 - \m} dr = \int_{u_2-iv_2}^{u_2+iv_2} \frac{\sqrt{(y+w_2)[(y-u_2)^2 + v_2^2]}}{2(y-y_+)(y-y_-)} dy, \\
&\;& \hskip -1cm \int_{u_1-iv_1}^{u_1+iv_1} \frac{ -\m a r} {(r^4 + a^2 r^2 - \m) \sqrt{r^6+ a^2 r^4 + \m a^2} } dr \nonumber \\
&\;& \hskip -1.5cm = \int_{u_2-iv_2}^{u_2+iv_2} \frac{ -\m a} {2(y-y_+)(y-y_-)\sqrt{(y+w_2)[(y-u_2)^2 + v_2^2]}} dy.
\eea
Here,  
\bea
&\;& \hskip -8cm u_2 = \frac{a^2}{6}(\lam^{1/3} + \lam^{-1/3} - 2), \\
&\;& \hskip -8cm v_2 = \frac{\sqrt{3} a^2}{6}(\lam^{1/3} - \lam^{-1/3}), \\
&\;& \hskip -8cm w_2 = \frac{a^2}{3}(\lam^{1/3} + \lam^{-1/3} + 1), \\
&\;& \hskip -8cm \lam = 1 + \frac{27\m}{2a^4} + 3\sqrt{3}\sqrt{\frac{\m}{a^4} + \frac{27\m^2}{4a^8}}, \\
&\;& \hskip -8cm  y_{\pm} = \frac{-a^2}{2} \pm \sqrt{\m + \frac{a^4}{4}}.
\eea
Similarly, define 
\bea
&\;& f_0(y_0,u_2,v_2,w_2) = \int_{u_2-iv_2}^{u_2+iv_2} \frac{\sqrt{(y+w_2)[(y-u_2)^2 + v_2^2]}}{y-y_0} dy, \\
&\;& f_m(y_0,u_2,v_2, w_2) = \int_{u_2-iv_2}^{u_2+iv_2} \frac{1}{(y-y_0)\sqrt{(y+w_2)[(y-u_2)^2 + v_2^2]}} dy.
\eea
We find 
\bea
&\;& \hskip -1cm f_0(y_0,u_2,v_2,w_2) = \frac{2}{3} \sqrt{u_2 + i v_2 + w_2} (2u_2 - w_2 - y_0)  \nonumber \\
&\;& \hskip 2.0cm \times \left\{ E\left[\sin^{-1} \left(  \sqrt{\frac{u_2 + i v_2 + w_2}{u_2 - i v_2 + w_2}}\, \right),\frac{u_2 - i v_2 + w_2}{u_2 + i v_2 + w_2} \right] - E\left[ \frac{u_2 - i v_2 + w_2}{u_2 + \i v_2 + w_2} \right] \right\} \nonumber \\
&\;& \hskip 1.8 cm + \frac{2}{3} \frac{1}{ \sqrt{u_2 + i v_2 + w_2} } [ 3y_0^2 -3(u_2 - iv_2 -w_2)y_0 +(2v_2^2 - 2i u_2 v_2 + i v_2 w_2 - 3 u_2 w_2)] \nonumber \\
&\;& \hskip 2.0cm \times \left\{ F\left[\sin^{-1} \left(  \sqrt{\frac{u_2 + i v_2 + w_2}{u_2 - i v_2 + w_2}}\, \right), \frac{u_2 - i v_2 + w_2}{u_2 + i v_2 + w_2} \right] - K\left[ \frac{u_2 - i v_2 + w_2}{u_2 + \i v_2 + w_2} \right] \right\} \nonumber \\
&\;& \hskip 1.8 cm - 2 \frac{1}{ \sqrt{u_2 + i v_2 + w_2} } [ (y_0 - u_2)^2 + v_2^2 ]\nonumber \\
&\;& \hskip 2.0cm \times \Biggl\{ \Pi\left[\frac{u_2 - i v_2 + w_2}{w_2 + y_0}, \sin^{-1} \left( \sqrt{\frac{u_2 + i v_2 + w_2}{u_2 - i v_2 + w_2}}\, \right), \frac{u_2 - i v_2 + w_2}{u_2 + i v_2 + w_2} \right]  \nonumber \\
&\;& \hskip 2.5cm - \Pi\left[\frac{u_2 - i v_2 + w_2}{w_2 + y_0}, \frac{u_2 - i v_2 + w_2}{u_2 + \i v_2 + w_2} \right] \Biggr\},
\eea
and
\bea
&\;& \hskip -1cm f_m(y_0,u_2,v_2,w_2) = \frac{2}{(w_2 + y_0) \sqrt{u_2 - i v_2 + w_2}} \nonumber \\
&\;& \hskip 2.0cm \times \left\{ F\left[\sin^{-1} \left(  \sqrt{\frac{u_2 - i v_2 + w_2}{u_2 + i v_2 + w_2}}\, \right),\frac{u_2 + i v_2 + w_2}{u_2 - i v_2 + w_2} \right] - K\left[ \frac{u_2 + i v_2 + w_2}{u_2 - \i v_2 + w_2} \right] \right\} \nonumber \\
&\;& \hskip 1.8 cm - \frac{2}{(w_2 + y_0) \sqrt{u_2 - i v_2 + w_2}} \nonumber \\
&\;& \hskip 2.0cm \times \Biggl\{ \Pi\left[\frac{w_2 + y_0}{u_2 - i v_2 + w_2}, \sin^{-1} \left( \sqrt{\frac{u_2 - i v_2 + w_2}{u_2 + i v_2 + w_2}}\, \right), \frac{u_2 + i v_2 + w_2}{u_2 - i v_2 + w_2} \right]  \nonumber \\
&\;& \hskip 2.5cm - \Pi\left[\frac{w_2 + y_0}{u_2 - i v_2 + w_2}, \frac{u_2 + i v_2 + w_2}{u_2 - \i v_2 + w_2} \right] \Biggr\}.
\eea
Comparing to the numerical integration, we find  
\bea
&\;& \hskip -4cm \d_0 = -2i\biggl\{  \frac{f_0(y_+,u_2,v_2, w_2) - f_0(y_-,u_2,v_2, w_2)}{y_+ - y_-} + i2\pi \left[ \frac{\sqrt{-\m y_-}}{2(y_+ - y_-)} \right] \biggr\}, \\
&\;& \hskip -4cm \d_m = -2i \biggl\{ \frac{-\m a[f_m(y_+,u_2,v_2,w_2) - f_m(y_-,u_2, v_2, w_2)]}{2(y_+ - y_-)} \biggr\}.
\eea
The magnitude of jump here is again determined by the residue of $q_0^{1/2}/\D$ at $r=r_+$.
The results are shown in Fig. 6 and 7.  There is no upper bound in $a$.  Again, there is a critical value $a_c \approx 0.516 \m^{1/4}$ and $\hat{\om}$ changes sign when $a>a_c$.
\vspace*{0.5cm}
\begin{figure}[tbh]
\centerline{\epsfxsize=8cm \epsffile{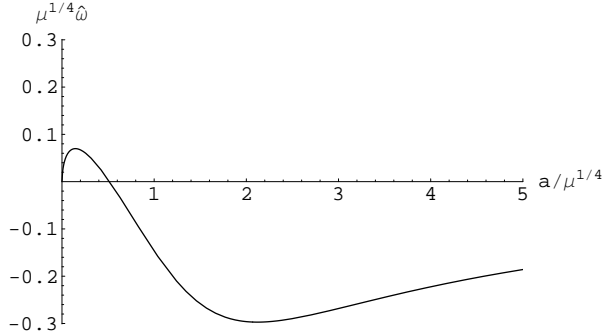}}
\caption{\label{fig:omega_hat_D=7} The analytic result for the real part of the highly damped QNM frequency $\hat{\om}(a)$ when $D=7$.}
\end{figure}

\vspace*{0.5cm}
\begin{figure}[tbh]
\centerline{\epsfxsize=8cm \epsffile{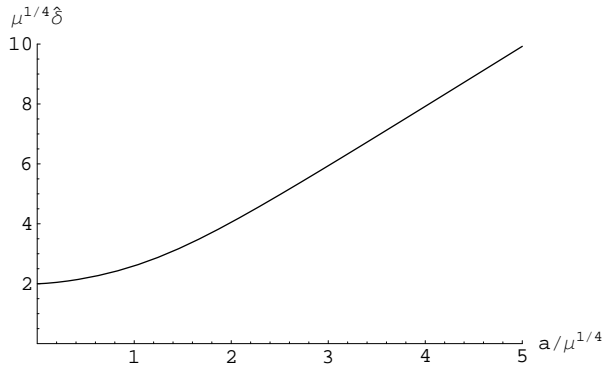}}
\caption{\label{fig:delta_hat_D=7} Analytic result for level spacing $\hat{\d}$ when $D=7$.}
\end{figure}

\section{Conclusion}

In sum, we have calculated the asymptotic form of quasinormal frequencies to leading order for Kerr black holes in four, five and seven dimensions. All the releveant quantities can be expressed in terms of elliptical integrals. In particular, the results are consistent with the numerical ones in four dimensions.  It would be interesting to generalize the method to other rotating black holes, especially Kerr-AdS black holes in five dimesions. It is pointed out in Ref. \cite{Horowitz} that a large black hole in AdS background corresponds to a thermal state in CFT. Therfore, studying the decay of the scalar field may tell us something about the decay of a perturbation of the thermal state \cite{AdSCFT}.  Extension to next to leading order is also desirable. One need to do systematic expansion in $c^{-1}$ in a way similar to that used in Ref. \cite{Brink}. Such calculation may allow us to do a more comprehesive comparison between the analytic and numerical results.

\section*{Acknowledgment} 
 
The author thanks Hing-Tong Cho for helpful discussions. The work is supported in part by the National Science Council and the National Center for Theoretical Sciences, Taiwan.

\end{document}